\documentclass{kapproc} 

\usepackage{procps} 

\usepackage[dvips]{graphicx}

\upperandlowercase

\kluwerbib 

\begin{document}



\articletitle{Collapse of the Primordial Gas Clouds in the Presence of UV Radiation Field}










\author{Jaroslaw Stasielak\altaffilmark{1}, Slawomir Stachniewicz\altaffilmark{2}, and
         Marek Kutschera\altaffilmark{2,1}}

\altaffiltext{1}{Institute of Physics, Jagiellonian University, ul. Reymonta 4,\\
30-059 Krakow, Poland}
\email{stasiela@th.if.uj.edu.pl}
\altaffiltext{2}{Astrophysics Division, H.Niewodnicza\'nski Institute of Nuclear Physics,\\ 
ul. Radzikowskiego 152, 31-342 Krakow, Poland}
\email{Slawomir.Stachniewicz@ifj.edu.pl}
\email{Marek.Kutschera@ifj.edu.pl}



\begin{abstract}
Our goal is to study the effects of the UV radiation from the first stars, quasars and hypothetical Super Heavy Dark Matter (SHDM) particle decays on the formation of primordial bound objects in the Universe. We trace the evolution of a spherically symmetric density perturbation in the Lambda Cold Dark Matter and MOND model, solving the frequency-dependent radiative transfer equation, non-equilibrium chemistry, and one-dimensional gas 
hydrodynamics. We concentrate on the destruction and formation processes of the $H_{2}$ 
molecule, which is the main coolant in the primordial objects.
\end{abstract}


\section{Introduction}
The existence of the first objects is a direct consequence of the growth of the primordial density fluctuations. At the beginning, there are linear density perturbations which expand with the overall Hubble flow. Subsequently, these perturbations can grow and form primordial clouds. Clouds with enough density contrast decouple from this flow and start to collapse. The kinetic energy of the infalling gas is dissipated through shocks and the cloud becomes pressure supported. The further evolution of the cloud is determined by its ability to cool sufficiently fast. Clouds which could not cool fast enough will stay in a pressure-supported stage and will not form any stars. The existence of the efficient cooling mechanism is necessary to continue the collapse of the cloud, its subsequent fragmentation and star formation. 

In our work we are interested in the first generation of stars which are still forming in the low mass clouds when first luminous objects already exist. These objects are made from the primordial gas so they are metals free. It is simply because the first stars did not have much time to produce them. These objects could be irradiated by the UV and X-rays radiation produced by the first stars, quasars and hypothetical Super Heavy Dark Matter (SHDM) particle decays (\cite{Do02}; \cite{Sh04}).

In the absence of metals, the most important cooling mechanism for low-mass primordial clouds is so called '$H_{2}$ cooling', i.e. cooling by radiation of excited rotational and vibrational states of $H_{2}$ molecule. The presence of initial mass fraction of the molecular hydrogen $H_{2}$ of only $10^{-6}$ is enough to trigger the final collapse of low mass clouds.

However, molecular hydrogen is fragile and can easily be photo-dissociated by photons with energies of 11.26 - 13.6 eV (Lyman and Werner bands)(\cite{Ms82}). Destruction of the $H_{2}$ would stop collapsing of the low mass clouds and decrease star formation rate. On the other hand X-rays radiation can increase production of the $H_{2}$ by enhancement of free electrons fraction. From above we see that the UV and X-rays radiation background alters the subsequent growth of cosmic structures. It regulates star formation rate so it has important implication to the re-ionization history of the Universe \mbox{(\cite{Ce03})}. It is therefore crucial to determine quantitatively the consequences of radiation feedback on the formation of early generation objects.

Feedback of the UV background to the collapse of the spherically symmetric primordial gas cloud was studied by many authors (\cite{Ta98}; \cite{Ki00}; \cite{Ki00b}; \cite{Ki00c}; \cite{Om01}). However their calculation was simplified. They have been using the so called self-shielding function (\cite{Dr96}). Our approach includes solving the frequency-dependent radiative transfer equation in the 'exact' way. We also try to check the evolution of the collapsing cloud in the MOND model.
 

\section{Description of the code}

In the Newtonian case dynamics is governed by the following equations:

\begin{eqnarray}
{dM \over dr} & = & 4\pi r^2 \varrho, \label{ciaglosc}\\
{dr \over dt} & = & v , \label{promien}\\
{dv \over dt} & = & -4\pi r^2 {dp \over dM}-{GM(r) \over r^2} ,
\label{predkosc}\\
{du \over dt} & = & {p \over \varrho^2} {d \varrho \over dt} + {\Lambda \over
\varrho} ,
\label{energia}
\end{eqnarray}

\noindent where $r$ is the radius of a sphere of mass $M$, $u$ is the internal
energy per unit mass, $p$ is the pressure and $\varrho$ is the mass
density. Here eq.(\ref{ciaglosc}) is the continuity equation, (\ref{promien})
and (\ref{predkosc}) give the acceleration and (\ref{energia}) accounts for
the energy conservation.
The last term in the eq.(\ref{energia}) describes cooling/heating
of the gas, with $\Lambda$ being the energy absorption (emission) rate per
unit volume.

$\Lambda$ consist of two parts, that is, the radiative cooling $\Lambda _{rad}$ and the chemical cooling $\Lambda_{chem}$. The former can be written as
\begin{equation}
	\Lambda_{rad} \left(m\right) = -\varrho \frac{\partial L\left(m\right)}{\partial m}. \label{cooling}
\end{equation}
\noindent The luminosity $L\left(m\right)$ is obtained by solving radiative transfer equation.

The net chemical cooling rate $\Lambda_{chem}$ is given by
\begin{equation}
\Lambda_{chem} = \varrho \frac{\partial \epsilon _{chem}}{\partial t}
\end{equation}
\noindent where $\epsilon _{chem}$ is the chemical binding energy per unit mass.

We use the equation of state of the perfect gas

\begin{equation} p= (\gamma -1) \varrho u , \end{equation}

\noindent where $\gamma = 5/3$, as the primordial baryonic matter after
recombination
is assumed to be composed mainly of monoatomic hydrogen and helium, with
the fraction of molecular hydrogen $H_2$ always less than $10^{-3}$.

In case of modified gravity (MOND) we do not include the dark component but it is 
necessary to modify equation (\ref{predkosc}). Details are shown in 
\cite{Sta05}.

In the simulations we have used the code described in \cite{Sta01}, based on
the codes described by \cite{Tho95} and \cite{Hai96}.
This is a standard, one-dimensional, second-order
accurate Lagrangian finite-difference scheme.

We start our calculations at $z=500$ or the end of the
radiation-dominated era. We use our own code to calculate the initial chemical 
composition and initial gas temperature. Initial overdensities may 
be calculated from the power spectrum which may 
be obtained e.g. using the {\sc CMBFAST} program by \cite{Sel96}.

We apply initial density profiles in the form of a single spherical
Fourier mode used also by \cite{Hai96}

\begin{equation} \varrho_b(r)=\Omega_b \varrho_c (1+\delta {\sin kr \over
kr})
,
\label{psinus} \end{equation}

\noindent where $\varrho_c$ is the critical density of the Universe,
$\varrho_c=3H^2/8\pi G$ with $H$ being the actual value of the Hubble
parameter. As the initial velocity we use the Hubble velocity:

\begin{equation}v(r)=Hr . \end{equation}

In our calculations we include nine species:
H, H$^-$, H$^+$, He, He$^+$, He$^{++}$, H$_2$, H$_2^+$ and e$^-$.
Full list of relevant chemical reactions, appropriate formulae, 
reaction, photoionization and photodissociation rates is given 
in \cite{Sta01}.


\section{Radiative transfer equation in spherical symmetry}

In addition to eq. (\ref{ciaglosc}) - (\ref{energia}) we solve non-relativistic, time-independent equation for radiation transport in spherical geometry:
\begin{equation}
	 \mu \frac{\partial I_{\nu}}{\partial r} + \frac{1 - \mu^{2}}{r} \frac{\partial I_{\nu}}{\partial \mu} = 
	\varrho \left\{ j_{\nu}\left(r\right)	-  \kappa_{\nu}\left(r\right) I_{\nu}		\right\} \textrm{,}
\end{equation}
where $I_{\nu} = I_{\nu} \left(r,\mu\right)$ is the intensity of radiation of frequency $\nu$, at radius $r$ and in the direction $\mu = \cos \theta$, where $\theta$ is the angle between the outward normal and photon direction. $j_{\nu}\left(r\right)$ is the emissivity and $\kappa_{\nu}\left(r\right)$ opacity at radius $r$.

Our scheme of solving this equation is similar to method from \cite{rad} and \cite{Yo80}. Detailed results will be described in a paper, which is in preparation.




\end{document}